# A FRW dark fluid with a non-linear inhomogeneous equation of state

*I. Brevik\*, E. Elizalde\*\*, O. Gorbunova\*\*\*, A.V. Timoshkin\*\*\**

*\*Department of energy and Process Engineering, Norwegian University of Science and Technology
N-7491 Trondheim, Norway
\*\* Instituto de Ciencias del Espacio (CSIC) and Institut d'Estudis Espacials de Catalunya (IEEC/CSIC)
Campus UAB, Facultat de Ciències, Torre C5-Parell-2a planta, E-08193 Bellaterra (Barcelona) Spain
\*\*\*Tomsk State Pedagogical University, Tomsk, Russia*

**Abstract**

A dark Friedman-Robertson-Walker fluid governed by a non-linear inhomogeneous equation of state is considered which can be viewed as a conveniently simple paradigm for a whole class of models which exhibit phase transitions from a non-phantom towards a phantom era (superacceleration transition). From another side, such dark fluid models may describe also quintessence-like cosmic acceleration. Thermodynamical considerations for the processes involved, which are of great importance in the characterization of the global evolution of the corresponding universe, are given too. Connecting the proposed equation of state with an anisotropic Kasner universe with viscosity, we are led to the plausible conjecture of a dark fluid origin of the anisotropies in the early universe.

# 1. Introduction

There is a lot of interest in the study of the nature of dark energy (for a review, see [1]), responsible for the acceleration of the cosmic expansion, which was initiated not far in the past and which is expected to continue to late times. Among the different possible models that have been considered in the literature, one that has a good probability to reflect (or at least to conveniently parameterize) what may be going on there, is a model in which dark energy is described by some rather complicated ideal fluid with an unusual equation of state (EoS). Very general dark fluid models can be described by means of an inhomogeneous equation of state [2]. Some particular examples of such kind of equations have been considered in Refs. [3]-[9], also including in some cases observational consequences of the corresponding generalized dark fluids [10]. Moreover, a dark energy fluid obeying a time-dependent equation of state [4,7] may be also successfully used with the purpose to mimic the classical string landscape picture [11], what is very interesting in order to establish a connection with a different fundamental approach. And, even more, it is known as well that a dark fluid satisfying a time-dependent equation of state can rather naturally lead to a phantom era [12], where the phantom field is able to mimic some features of the underlying quantum field theory [13].

In this paper we will investigate a specific model for a dark fluid with a non-linear equation of state, which seems to be particularly natural and exhibits quite nice properties. In this sense it can be considered as a conveniently simple paradigm for that class of models. We start by considering, in Sect. 2, an inhomogeneous equation of state for the universe and find its solutions in terms of the Hubble constant and the scale factor, investigating the transitions from the non-phantom to the phantom era, in particular a transition towards superacceleration, that is, the case when the third derivative of the scale factor is also positive. Thermodynamical considerations for the processes involved ---which are of great importance for the characterization of the global evolution of our universe--- are provided in Sect. 3. In Sect. 4 (which is not directly related to the previous sections) we provide a connection of a specific class of the models in this paper with the Kasner metric, as applied to a viscous cosmic fluid. Finally, the last section is devoted to a summary and conclusions.

# 2. Inhomogeneous equation of state for the universe and its solution

We assume that our universe can be conveniently described by a field giving rise to an ideal fluid (dark energy) which obeys a non-linear inhomogeneous equation of state depending on time:

$$p = w(t)\rho + f(\rho) + \Lambda(t), \qquad (1)$$

where $w(t)$ and $\Lambda(t)$ depend on time, $t$, and where $f(\rho)$ is an arbitrary function, in the general case. This equation of state, for the case when $\Lambda(t) = 0$, was examined in Refs. [2,4,7] for different functions $w(t)$ and $\Lambda(t)$. Generally speaking, an effective equation of state of this class is typical in modified gravity (see [14] for a review).

Let us write down the energy conservation law, which is

$$\dot\rho + 3H(\rho + p) = 0 \qquad (2)$$

and the corresponding FRW equation for a spatially flat FRW universe, namely

$$\frac{3}{\chi^2}H^2 = \rho. \qquad (3)$$

Using Eqs. (1) and (3), we get

$$\dot{\rho}+\sqrt{3}\chi\rho^{1/2}[(w(t)+1)\rho+f(\rho)+\Lambda(t)]=0, \qquad (4)$$

where $\rho$ is the energy density, $p$ the pressure, $H=\dfrac{\dot{a}}{a}$ the Hubble parameter, $a(t)$ the scale factor of the three-dimensional flat Friedman universe, and $\chi$ the gravitational constant. It is known that, with quite good precision, an ideal fluid model with an inhomogeneous equation of state can satisfy the very recent observational data [10].

In papers [4,5] a model was considered in which both the functions $w(t)$ and $\Lambda(t)$ were periodic. We assume here, for simplicity, that the function $\Lambda(t)=0$ and choose the parameter $w(t)$ to be linearly depended of time, that is,

$$w(t) = a_1 t + b$$

(what is a very reasonable choice in view of the latest observational proposals at a determination of the evolution of the effective equation of state parameter), while for the function $f(\rho)$ we choose it (as in Ref. [9]) to be: $f(\rho) = A \cdot \rho^\alpha$, where $A, \alpha$ are constants. Then, Eq. (4) acquires the following form

$$\dot{\rho}+\sqrt{3}\chi\rho^{3/2}(a_1 t + b + 1)+\sqrt{3}\chi A \rho^{\frac{1}{2}+\alpha} = 0. \qquad (5)$$

We shall investigate, for further simplicity, the case when $\alpha = \dfrac{1}{2}$. The solution of Eq. (5) looks as

$$\rho(t) = \dfrac{1}{\left[S\cdot e^{\frac{\sqrt{3}}{2}A\chi t} - \dfrac{1}{A}(a_1 t + b + 1 + \dfrac{\sqrt{3}}{2}\dfrac{a_1\chi}{A})\right]^2}, \qquad (6)$$

and the Hubble parameter is

$$H(t) = \dfrac{\chi}{\sqrt{3}\left[S\cdot e^{\frac{\sqrt{3}}{2}A\chi t} - \dfrac{1}{A}(a_1 t + b + 1 + \dfrac{\sqrt{3}}{2}\dfrac{a_1\chi}{A})\right]}, \qquad (7)$$

where $S$ is an integration constant. It is quite clear that for some specific values of parameters where the effective equation of state parameter is around -1, the Friedman universe expands with acceleration.
The time derivative of $H(t)$ becomes

$$\dot{H}(t) = \dfrac{\dfrac{a_1}{A}\dfrac{\chi}{\sqrt{3}} - S\dfrac{\chi^2}{2}A\cdot e^{\frac{\sqrt{3}}{2}\chi A t}}{\left[S\cdot e^{\frac{\sqrt{3}}{2}A\chi t} - \dfrac{1}{A}(a_1 t + b + 1 + \dfrac{2a_1}{\sqrt{3}A\chi})\right]^2}. \qquad (8)$$

This time derivative is equal to zero when $t_0 = \dfrac{2}{\sqrt{3}\chi A}\ln\dfrac{2a_1}{\sqrt{3}S\chi A^2}$.

If $a_1 > 0$, $A > 0$, and $t < t_0$, then $\dot{H} > 0$ that is, the universe is accelerating, and if $t > t_0$, one gets $\dot{H} < 0$, and correspondingly a decelerating universe. There is here a transition from a phantom epoch to a non-phantom one. Indeed, at the moment when the universe passes from the phantom to the non-phantom era, the Hubble parameter is equal to

$$H_m = \frac{A^2\chi^2}{2a_1\left[1-\left(\ln\frac{2a_1}{\sqrt{3}S\chi A^2}+\frac{\sqrt{3}\chi A^2(b+1)}{2a_1}+\frac{3}{4}\chi^2\right)\right]}. \qquad (9)$$

In the phantom phase, $\dot{\rho}>0$, the energy density grows and the Universe is expanding. In the non-phantom phase, $\dot{\rho}<0$ and the energy density decreases.

If $t\to+\infty$, then $H(t)$ and $\rho(t)\to 0$, so that the phantom energy decreases. The cosmology singularity does not appear in this case. A plot of the function $H$ versus time $t$ is shown in Fig. 1:

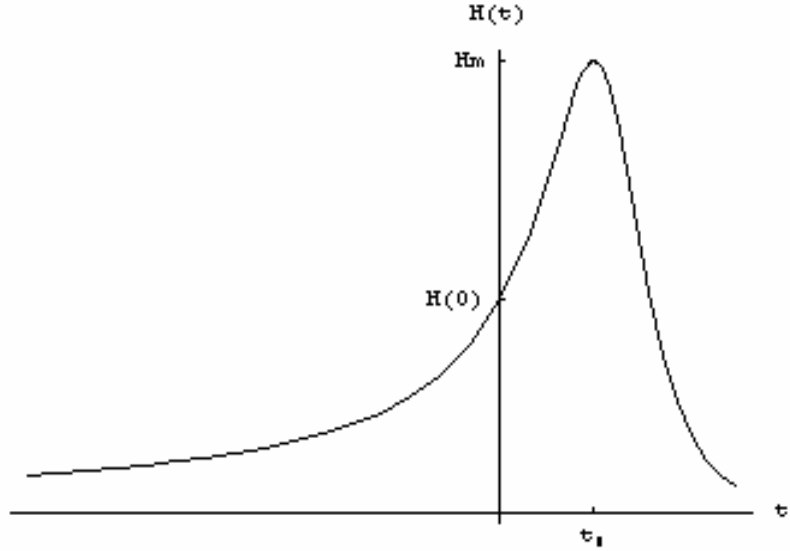

Fig. 1. The Hubble parameter $H$ as a function of time, for the case $a_1 > 0$.

The value of $H(t)$ at $t=0$ is given by

$$H(0) = \frac{\chi}{\sqrt{3}\left[S-\frac{1}{A}\left(b+1+\frac{\sqrt{3}}{2}\frac{a_1\chi}{A}\right)\right]},$$

and the scale factor in this case, with $a_1 > 0$, takes the following form

$$a(t) = e^{\int H(t)\,dt} = e^{\int \frac{\chi\,dt}{\sqrt{3}\left[S\cdot e^{\frac{\sqrt{3}}{2}A\chi t}-\frac{1}{A}(a_1 t+b+1+\frac{\sqrt{3}}{2}\frac{a_1\chi}{A})\right]}}. \qquad (10)$$

Let us assume now, as a second case, that $a_1 = 0$. Then, the equation of state acquires the form

$$p(t) = b\rho(t) + A\rho^{\frac{1}{2}}. \qquad (11)$$

The Hubble's parameter is

$$H(t) = \frac{1}{e^{\frac{\sqrt{3}}{2}A\chi t} - \frac{b+1}{A}}, \quad (12)$$

and the derivatives of $H(t)$ and $\rho(t)$ correspondingly become

$$\dot{H}(t) = -\frac{\frac{\sqrt{3}}{2}A\chi e^{\frac{\sqrt{3}}{2}A\chi t}}{\left(e^{\frac{\sqrt{3}}{2}A\chi t} - \frac{b+1}{A}\right)^2}, \quad (13)$$

$$\dot{\rho}(t) = -\frac{3\sqrt{3}\frac{A}{\chi}e^{\frac{\sqrt{3}}{2}A\chi t}}{\left(e^{\frac{\sqrt{3}}{2}A\chi t} - \frac{b+1}{A}\right)^3}. \quad (14)$$

When $t_0 = \frac{2}{\sqrt{3}\chi A}\ln\frac{b+1}{A}$ a future cosmological singularities appear [12]. Note that such singularities can be of very different types [15]. The scale factor is given by the expression

$$a(t) = e^{-\frac{A}{b+1}t}\left(e^{\frac{\sqrt{3}}{2}A\chi t} - \frac{b+1}{A}\right)^{\frac{2}{\sqrt{3}\chi(b+1)}}. \quad (15)$$

The derivative of the scale factor is

$$\dot{a}(t) = e^{-\frac{A}{b+1}t}\left(e^{\frac{\sqrt{3}}{2}A\chi t} - \frac{b+1}{A}\right)^{\frac{2}{\sqrt{3}\chi(b+1)}-1}, \quad (16)$$

and we have that $\dot{a}(t) = 0$, for $t = t_1 = \frac{2}{\sqrt{3}\chi A}\ln\frac{b+1}{A}$, while the second derivative of scale factor is

$$\ddot{a}(t) = \dot{a}^2 e^{\frac{A}{b+1}t}\left(1 - \frac{\sqrt{3}}{2}A\chi e^{\frac{\sqrt{3}}{2}A\chi t}\right), \quad (17)$$

and we have also that $\ddot{a}(t) = 0$, for $t_2 = \frac{2}{\sqrt{3}\chi A}\ln\frac{2}{\sqrt{3}\chi A}$.

Provided $A > 0$ and $b > -1$, for values of $t < t_2$ it turns out that the first and the second derivatives of the scale factor are both positive (i.e., the universe expands with acceleration), while for $t > t_2$ the first derivative is positive but the second derivative is negative (in this case it is still expanding but the expansion is decelerated). This is a very interesting transition.

A more detailed analysis can be carried out that involves the third time-derivative of the scale factor. This is given by the following, rather involved expression

$$\dddot{a}(t) = A\ddot{a}(t)\left[-\frac{1}{b+1} + \frac{\sqrt{3}\chi e^{\frac{\sqrt{3}}{2}A\chi t}}{e^{\frac{\sqrt{3}}{2}A\chi t} - \frac{b+1}{A}}\left(\frac{1}{\sqrt{3}\chi(b+1)} - 1\right) + \frac{3}{4}A\chi^2 e^{\frac{\sqrt{3}}{2}A\chi t}\left(\frac{\sqrt{3}}{2}A\chi e^{\frac{\sqrt{3}}{2}A\chi t} - 1\right)^{-1}\right]. \quad (18)$$

A positive (resp. negative) third derivative signalizes the rate at which (super)acceleration takes place. In an expanding universe with an accelerated expansion (as is the case of our own, at present), a positive value of the third derivative will clearly point towards a superacceleration (phantom) phase. In our model, a cumbersome but rather straigthforward calculation leads to the conclusion that this is actually the case. The situation is rather remarkable: it turns out that for $t_1 < t < t_2$ the universe suffers a (super)accelerated expansion with a positive acceleration rate (positive third derivative), while for $t > t_2$ its expansion is not only decelerating, but the deceleration rate diminishes too. For earlier times, $t < t_1$, the universe expansion is accelerating with a diminishing rate, before going through the first transition at $t = t_1$. This is summarized in Table 1. Hence, within this so simple model, the (super)acceleration regime may well describe the currently observable universe.

|  | $\left(-\infty, \frac{2}{\sqrt{3}A\chi} \ln \frac{b+1}{A}\right)$ | $\left(\frac{2}{\sqrt{3}A\chi} \ln \frac{b+1}{A}, \frac{2}{\sqrt{3}A\chi} \ln \frac{2}{\sqrt{3}A\chi}\right)$ | $\left(\frac{2}{\sqrt{3}A\chi} \ln \frac{2}{\sqrt{3}A\chi}, +\infty\right)$ |
|---|---|---|---|
| $\dot{a}(t)$ | >0 | >0 | >0 |
| $\ddot{a}(t)$ | >0 | >0 | <0 |
| $\dddot{a}(t)$ | <0 acceleration | >0 superacceleration | <0 deceleration |

*Table. 1. Complete description of the evolutionary transitions according to the values of the scale factor $a(t)$ and its first, second and third derivatives, which characterize two types of accelerating (for $\ddot{a}(t) > 0$) and one of decelerating (for $\ddot{a}(t) < 0$) expansions of the universe.*

Let us finally consider the case of the non-linear inhomogeneous equation of state $\Lambda(t) = ct + d$, where $c, d$ are arbitrary constants. The equation of motion is

$$\dot{\rho} + \sqrt{3}\chi\rho^{1/2}\left[(a_1 t + b)\rho + A\rho^{\frac{1}{2}} + ct + d\right] = 0 \qquad (19)$$

and the solutions look like

$$\rho(t) = \left(\frac{3A_1}{\chi\left[Be^{C(t+D)^2}erf(t) - 1\right]}\right)^2. \qquad (20)$$

The Hubble's parameter is given by the expression

$$H(t) = \frac{A_1}{Be^{C(t+D)^2}erf(t) - 1}, \qquad (21)$$

where $erf(t)$ is the probability integral. The derivative of $H(t)$ is equal to

$$\dot{H}(t) = -\frac{2A_1 Be^{C(t+D)^2}}{\left[B \cdot e^{C(t+D)^2}erf(t) - 1\right]^2} \cdot \left[C(t+D) \cdot erf(t) + \frac{1}{\sqrt{\pi}} \cdot e^{-C(t+D)^2}\right], \qquad (22)$$

where $A_1 = \dfrac{\chi}{\sqrt{3}} \cdot \dfrac{C}{a_1}$, $B = \sqrt{\dfrac{2\pi}{\sqrt{3}\chi C}} \dfrac{\sqrt{3}\chi}{4} \left[ \dfrac{a_1 C}{b+1} - (d+A) \right]$, $C = \dfrac{\sqrt{3}\chi}{4} c$, $D = \dfrac{d+A}{c}$.

At $t = t_2$, where $t_2$ is the solution of the equation $erf(t) = \dfrac{1}{B} e^{-C(t+D)}$, a future cosmological singularity occurs. There, the energy density and the Hubble parameter simultaneously approach infinity. This is depicted in Fig. 2. Hence, such a model actually describes a phantom era.

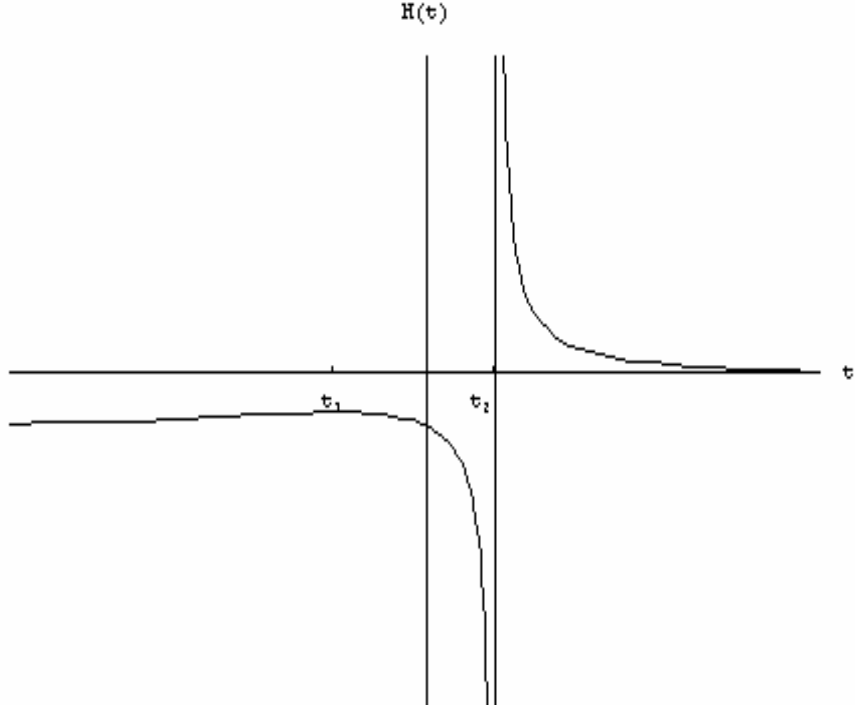

*Fig. 2. The future cosmological singularity which appears at $t = t_2$, where both the energy density and the Hubble parameter simultaneously tend to infinity.*

This finishes the investigation of the different acceleration regimes one can get from the postulated non-linear inhomogeneous EoS ideal fluid.

### 3. Some thermodynamical considerations.

We will consider in this section the main thermodynamical relation:

$$TdS = dE + \delta A. \qquad (23)$$

For the simple system, $\delta A = pdV$, then:

$$TdS = dE + pdV = (p + \rho)dV + Vd\rho. \qquad (24)$$

The volume of the system is: $V = a^3(t) \cdot \int \sqrt{\gamma} d^3 x$. Let us write the equation of state under the form

$$p = (a_1 t + b)\rho + A\rho^{1/2} + \Lambda(t), \tag{25}$$

where $\Lambda(t) = ct + d$. Then,

$$TdS = \left[(a_1 t + b + 1)\rho + A\rho^{\frac{1}{2}} + \Lambda(t)\right]dV + Vd\rho. \tag{26}$$

As before, we consider the particular equation of state with $\Lambda(t) = 0$. Then, the entropy multiplied by the temperature is equal to

$$TdS = -a^3(t) \cdot \frac{\frac{a_1}{A}\left(\frac{3}{2}\chi + 2\right)}{\left[Se^{\frac{\sqrt{3}}{2}A\chi t} - \frac{1}{A}\left(a_1 t + b + 1 + \frac{\sqrt{3}}{2}\frac{a_1\chi}{A}\right)\right]^3}\left(\int d^3x\sqrt{\gamma}\right)dt. \tag{27}$$

Assume now that $a_1 = 0$, then $p(t) = b\rho(t) + A\rho^{\frac{1}{2}}(t)$ and

$$TdS = 3a^2(t)\left(\int d^3 x\sqrt{\gamma}dt\right) \cdot e^{-\frac{A}{b+1}t}\left[\left(e^{\frac{\sqrt{3}}{2}A\chi t} - \frac{b+1}{A}\right)^{\frac{1}{\sqrt{3}\chi(b+1)}-1}\right]^2$$

$$\frac{\frac{\sqrt{3}A}{\chi}\left(e^{\frac{\sqrt{3}}{2}A\chi t}\right)^2 - (b+1)(1+\frac{\sqrt{3}}{\chi})e^{\frac{\sqrt{3}}{2}A\chi t} + \frac{(b+1)^2}{A} + A}{\frac{b+1}{A} - e^{\frac{\sqrt{3}}{2}A\chi t}}. \tag{28}$$

If $A > 0$ and $b > -1$, then $TdS > 0$, the temperature and the entropy have the same sign. (Note that in the phantom era it often happens that the entropy is negative [3]). Similar considerations can be extensible to other kind of specific ideal fluids.

### 4. The Kasner universe as a model

Given the expression (1) as a fundamental ansatz of our paper, it would seem physically desirable to connect this relation to the equation of state as derived from a specific universe model (in other words, some physical motivation for a specific EoS might be required). As we shall see, some insight in this direction can be obtained by drawing into consideration the Kasner universe as an example. Note that this may seem to be quite disconnected from the previous material, but the relation will be much more clear at the end. Crucial in this context is the fact that the Kasner metric ---known in general to possess anisotropy properties--- is to be applied to a cosmic fluid which is viscous. From ordinary fluid mechanics we know that the shear viscosity η comes into play whenever there are fluid sheets sliding with respect to each other. In connection with the Kasner metric, the occurrence of a shear viscosity thus appears to be most natural. Moreover we shall, to begin with, assume that there is also a bulk viscosity, ζ, present. Bulk viscosity is a concept that is compatible with the cosmic fluid being spatially isotropic. Our exposition follows the lines presented earlier in [3] and [18].

The Kasner metric is of the form

$$ds^2 = -dt^2 + t^{2p_1}dx^2 + t^{2p_2}dy^2 + t^{2p_3}dz^2, \tag{29}$$

where the numbers $p_1, p_2, p_3$ are constants. Let us define two new numbers S and Q by

$$S = \sum_{i=1}^{3} p_i, \qquad Q = \sum_{i=1}^{3} p_i^2 \ . \tag{30}$$

In a vacuum,

$$S = Q = 1 \ . \tag{31}$$

The (00)-component of Einstein's equation yields, with $\chi^2 = 8\pi G$,

$$S - Q + \frac{3}{2}\chi^2 t \varsigma S = \frac{1}{2}\chi^2 t^2 (\rho + 3p) \tag{32}$$

(the thermodynamic pressure $p$ is not to be confused with the Kasner parameters), and in the spatial directions $i = 1, 2, 3$

$$p_i \left(1 - S - 2\chi^2 t\eta\right) + \frac{1}{2}\chi^2 t \left(\varsigma + \frac{4\eta}{3}\right) S = -\frac{1}{2}\chi^2 t^2 (\rho - p) \ . \tag{33}$$

Here, both viscosity coefficients appear. This kind of formalism means physically that we work to first order in the deviation from thermodynamic equilibrium.

Let us now recall the usual kind of reasoning: one assumes that the viscosity coefficients η and ζ are given. Then, the single equation (32), plus the three equations (33), together with the extra equation of state, gives in all five equations. These are enough to determine the five unknowns ($p_1, p_2, p_3, \rho, p$). As shown in [18], one can write the time dependences as

$$\rho(t) = \rho_0 t^{-2}, \quad p(t) = p_0 t^{-2}, \quad \varsigma(t) = \varsigma_0 t^{-1}, \quad \eta(t) = \eta_0 t^{-1}, \tag{34}$$

where ($\rho_0, p_0, \eta_0, \varsigma_0$) are constants. This means that $t = 0$ is taken as the initial singular point.

It is however possible to argue in a different way. Assume that the Kasner parameters ($p_1, p_2, p_3$) are given initially. This means that $S$ and $Q$ are known. We require now that the shear viscosity $\eta$ is such that it satisfies the equation

$$1 - S = 2\chi^2 t\eta \ . \tag{35}$$

We have then three equations at our disposal: equation (32) as before, equations (33) which now reduce to one single equation,

$$\left(\varsigma + \frac{4\eta}{3}\right) S = -t(\rho - p), \tag{36}$$

and equation (35). These are sufficient to determine the three unknowns $(\rho, p, \eta)$. The bulk viscosity $\varsigma$ is uninfluenced by this procedure, and is merely an arbitrary input parameter.

From equation (35), we eliminate $t$:

$$t = \frac{1-S}{2\chi^2 \eta}, \tag{37}$$

and insert it into equation (36), to get

$$p_0 = \rho_0 + \left(\varsigma_0 + \frac{4\eta_0}{3}\right)\left(1 - 2\chi^2 \eta_0\right), \tag{38}$$

where the scaling (34) has been used. This equation can be looked upon as some form of equation of state. It is not imposed from outside, however, but is a consequence of Einstein's equations, together with the condition (35).

Eq. (38) is actually of the same inhomogeneous form as in the recent work [7]. It corresponds to a maximally stiff (Zeldovich) fluid in the sense that the velocity of sound is equal to $c$. Usually, when the shear viscosity is small,

$$2\chi^2 \eta_0 < 1, \tag{39}$$

one has $p_0 > \rho_0$ so that the dominant energy condition is broken.

But if $2\chi^2 \eta_0$ is large, it is possible to be in the phantom region where $p_0 < -\rho_0$. Moreover, it seems convenient to set the bulk viscosity $\varsigma$ equal to zero (this is also motivated by the fact that in conformal field theories, one must have $\varsigma = 0$ in order to satisfy the traceless condition on the energy-momentum tensor). Then, the equation of state simply becomes

$$p_0 = \rho_0 + \frac{4\eta_0}{3}(1 - 2\chi^2 \eta_0). \tag{40}$$

When the equation is written in terms of time-dependent quantities,

$$p(t) = \rho(t) + \frac{4\eta_0}{3t^2}(1 - 2\chi^2 \eta_0), \tag{41}$$

we see that it has actually the same form as expression (1), with

$$w(t) = 1, \quad f(\rho) = 0, \quad \Lambda(t) = \frac{4\eta_0}{3t^2}(1 - 2\chi^2 \eta_0). \tag{42}$$

The only physical restriction here, as following from the second law of thermodynamics, is that $\eta_0$ must be positive or zero. In this way, we see that the Kasner universe provides a physical link for our fundamental ansatz (1). There appears some interesting possibility here, which will be discussed in more detail elsewhere, that the dark energy fluid EoS be related with anisotropies in the early time universe.

## 5. Summary and conclusions

We can summarize the results obtained above for the model we have considered here by means of Table 1, corresponding to the equation of state under study, namely $p = b\rho + A\rho^{\frac{1}{2}}$, with $0 < A < \frac{2}{\sqrt{3}\chi}$ and $b > -1$. We have been able to give a complete description of the evolutionary transitions according to the values of the scale factor $a(t)$ and its first, second and third derivatives, which characterize different types of accelerating (for $\ddot{a}(t) > 0$) and decelerating (for $\ddot{a}(t) < 0$) expansions of the universe (only the case with $\dot{a}(t) > 0$ was considered). Accordingly, we could distinguish three regions where the behavior is different, two of them corresponding to accelerated expansion, where the rate of acceleration decreases for primordial times (normal acceleration) and then, after the corresponding transitions, starts to increase (supperaccelerating case), and a third region (after the singularity), where we still have expansion but with deceleration with a negative rate. It is intriguing that such a simple model could give rise to such different behaviors. Adding a cosmological constant like term to the equation of state led us to the conclusion that at some $t = t_2$ a future singularity forms, in the sense that the energy density and the Hubble parameter simultaneously approach infinity at this value. Some thermodynamical considerations have also been given. To be remarked is the fact that not so many ingredients go into this simple model, however, the results are quite promising, surely based in its non-linearity.

Here a remark is in order. We have considered a pure dark energy model. However, it is easy to introduce into our considerations usual matter; in that case, one can assume that dark energy appears suddenly at some point at the end of the deceleration phase. And under these conditions, before the dark energy epoch a matter dominated phase may be easily constructed. It is not difficult to study such a model analytically, the difference being that the equation of state consists of several components: dark matter, usual matter, and dark energy (of the type we have considered).

It was really challenging to try to derive a phenomenological model of this sort directly from a fundamental theory. A first approach towards this goal was given in Sect. 4 of this paper, where the class of models here considered could indeed be connected with the equation of state as derived from a specific universe model, namely the well known anisotropic Kasner universe, with viscosity, that was there seen to provide a physical link for our starting ansatz (1).

Still, it would be interesting to consider the reconstruction method [16] for this dark fluid and to represent the reconstructed model in terms of the scalar-tensor theory potentials [17], a quite well established procedure nowadays. We hope to be able to carry out this program in a subsequent publication.

*Acknowledgments.*

We thank Sergei Odintsov for very valuable discussions. OG and AVT have been supported by a Grant of the Scientific School, LRSS № 4489.2006.02., and by Grant RFBR 06-01-00609. EE was supported in part by MEC (Spain), projects BFM2006-02842, PR2006-0145, FIS2005-25313-E, and by AGAUR (Generalitat de Catalunya), contract 2005SGR-00790.

REFERENCES

1. E. Copeland, M. Sami , S. Tsujikawa , Int. Mod. Phys, D 15, 1753 (2006); T. Padmanabhan , Phys .Repts. 380, 235 (2003); L. Perivolaropoulos, astro-ph/ 0601014; I. Ya Arefeva and I. Volovich, hep-th/ 0612098.
2. S. Nojiri and S. D. Odintsov, Phys. Rev. D 72 , 023003 (2005) [hep-th/ 0505215]; S. Nojiri and S. D. Odintsov Phys. Lett. B 639, 144 (2006) [hep-th /0606025].
3. I. Brevik , S. Nojiri, S. D. Odintsov, L. Vanzo, Phys. Rev. D 70, 043520 (2004) [hep-th/ 0401073].
4. S. Nojiri, S. D. Odintsov, Phys. Lett. B 637, 139 (2006) [hep-th/ 0603062].
5. O. Gorbunova, hep-th/ 0608043.
6. I. Brevik, O. Gorbunova , Gen. Rel. Grav. 37, 2039 (2005) [gr-qc/ 0504001].
7. I. Brevik, O. G. Gorbunova, A. V. Timoshkin, European Phys. J. C 51, 179 (2007) [gr-qc/0701089].
8. J. Ren, X. H. Meng, astro-ph/ 0605010; J. Ren, X. H. Meng, Phys. Lett. B 636, 5 (2006); M. Hu, X. H. Meng, Phys. Lett. B 635, 186 (2006); A. Balaguera-Antonez, M. Nowakowski, arXiv: 0704.1871; V. Cardone, C.Tortora, A. Troisi, S. Capozziello, Phys. Rev. D 73, 043508 (2006).
9. S. Nojiri, S. D. Odintsov, Phys. Rev. D 70, 103522 (2004) [hep-th/0508170]; E. Elizalde, S. Nojiri, S. D. Odintsov, P. Wang, Phys. Rev., D 71, 103504 (2005) [hep-th/0502082]; S. Capozziello, S. Nojiri, S. D. Odintsov, Phys. Lett. B 634, 93 (2006) [hep-th/0512118].
10. S. Capozziello, V. Cardone, E. Elizalde, S. Nojiri, S. D. Odintsov, Phys. Rev. D 73, 043512 (2006) [astro-ph/0508350].
11. S. Nojiri, S. D. Odintsov, hep-th/0702031.
12. B. McInnes, JHEP 0208, 029 (2002).
13. S. Nojiri, S. D. Odintsov, Phys. Lett. B 562 , 147 (2003) [hep-th/ 0303117].
14. S. Nojiri, S. D. Odintsov, Int. J. Geom. Meth. Mod. Phys. 4, 115 (2007) [hep-th/0601213]; S. Nojiri, S. D. Odintsov, Phys. Rev. D 68 , 123512 (2003) [hep-th/0307288).
15. S. Nojiri, S. D. Odintsov, S. Tsujikawa, Phys. Rev. D 71, 063004 (2005) [hep-th/0501025].
16. S. Nojiri, S. D. Odintsov, hep-th/0611071.
17. S. Nojiri, S. D. Odintsov, H. Stefancic, Phys. Rev. D 74, 023003 (2006) [hep-th/0505215]; E. Elizalde, S. Nojiri, S. D. Odintsov, Phys. Rev. D 70, 043539 (2004) [hep-th/0405134].
18. I. Brevik, S. V. Pettersen, Phys. Rev. D 61, 127305 (2000).